\documentclass[prl,twocolumn,showpacs,preprintnumbers,amsmath,amssymb]{revtex4}
\usepackage{graphicx}
\usepackage{color}
\usepackage{amsmath}
\usepackage{epstopdf}
\usepackage{amsbsy}
\usepackage{dcolumn}
\usepackage{bm}

\pdfoutput=1
\begin{document}

\title{Nonequilibrium transport through a spinful quantum dot with superconducting leads}
\author{B. M. Andersen$^1$, K. Flensberg$^1$, V. Koerting$^2$, J. Paaske$^1$}
\affiliation{
$^1$Niels Bohr Institute, University of Copenhagen, DK-2100 Copenhagen \O, Denmark\\
$^2$NBIA, Niels Bohr Institute, University of Copenhagen, Blegdamsvej 17, DK-2100 Copenhagen \O, Denmark}
\date{\today{}}

\begin{abstract}
We study the nonlinear cotunneling current through a spinful quantum dot contacted by two superconducting leads. Applying a general nonequilibrium Green function formalism to an effective Kondo model, we study the rich variation in the IV-characteristics with varying asymmetry in the tunnel coupling to source and drain electrodes. The current is found to be carried respectively by multiple Andreev reflections in the symmetric limit, and by spin-induced Yu-Shiba-Russinov bound states in the strongly asymmetric limit. The interplay between these two mechanisms leads to qualitatively different IV-characteristics in the cross-over regime of intermediate symmetry, consistent with recent experimental observations of negative differential conductance and re-positioned conductance peaks in sub-gap cotunneling spectroscopy.
\end{abstract}

\pacs{73.63.Kv, 73.23.Hk, 74.45.+c, 74.55.+v}
\maketitle

During the last decade, a number of experiments have explored the problem of coherent electron transport through tunnel junctions and quantum dots coupled to superconducting leads\cite{Sheer:1997, Buitelaar:2002, Buitelaar:2003, Kasumov:2005, JarilloHerrero:2006, Heersche:2006, Xiang:2006} and the observation of sub-gap structure with steps near $eV=2\Delta/n$ ($n=1,2,3,\ldots$) in the low-bias current-voltage (IV) characteristics has been rationalized in terms of multiple Andreev reflections (MAR)\cite{Averin95, Bratus:1995, Cuevas96}. Interestingly, recent experiments have revealed that odd-occupied (Coulomb blockaded) quantum dots exhibit novel transport characteristics not included within standard MAR-mediated transport\cite{Buitelaar:2003, Eichler:2007, Jespersen:2007, Buizert:2007, Grove:2009, Pillet:2010}.

The physics of odd-occupied quantum dots coupled to superconducting leads is intimately tied to the study of magnetic impurities in superconductors. Magnetic impurities are known to generate localized Yu-Shiba-Rusinov (YSR) bound states inside the superconducting gap, offset from the gap edge roughly by the magnitude of the exchange coupling\cite{Balatsky:2006}. These bound states serve as informative signatures in tunneling spectroscopy of superconductor surfaces\cite{Balatsky:2006,Moca:2008} and very recently they have also been probed by scanning tunneling spectroscopy of MnPc molecules adsorbed on Pb(111) surfaces\cite{Pascual:2011}. The question remains, however, which signatures are to be expected in a generic voltage-biased quantum dot transport experiment, where the system is driven out of equilibrium by a finite Josephson frequency $2eV/\hbar$.

Recent theoretical work studied Andreev bound states induced by a spinful quantum dot coupled to a superconductor, with focus on the competition between Kondo and superconducting singlet correlations\cite{Glazman:1989, Satori:1992, Rozhkov:1999, Clerk:2000, Yoshioka:2000, Vecino:2003, Choi:2004, Siano:2004, Oguri:2004, Sellier:2005, Bauer:2007, Karrasch:2008, Hecht:2008, Meng:2009}. From detailed calculations of low-energy spectral properties, one may thus infer about e.g. the Josephson current carried by bound states. Far less is known about nonlinear transport through an interacting, voltage-biased quantum dot, the focus of the present paper\cite{Avishai:2001, LevyYeyati:2003, Liu:2004, DellAnna:2008}.

\begin{figure}[hb!]
\includegraphics[clip=true,width=0.9\columnwidth]{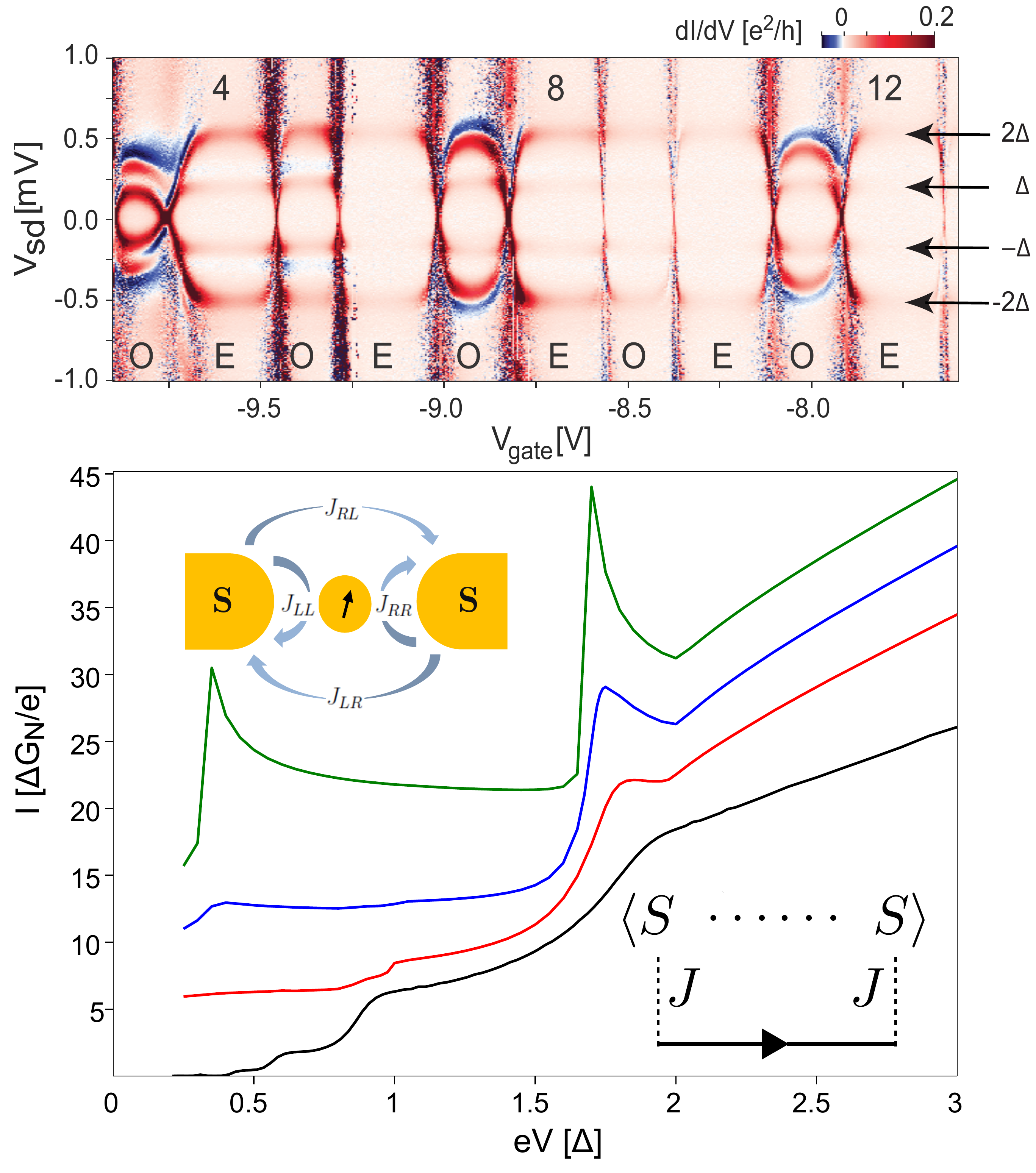}
\caption{(Color online) The top figure shows the conductance versus gate ($V_{\mbox{gate}}$) and bias ($V_{\mbox{sd}}$) voltage for a carbon nanotube quantum dot coupled to superconducting leads revealing a qualitatively different sub-gap structure at fillings 3, 7, and 11. For more details on the experimental setup see Ref. \cite{Grove:2009}. The bottom figure shows calculated IV characteristics through a S/QD/S cotunnel junction obtained within the present approach. For all curves $J_{LL}/\Delta=11.9$ and (top to bottom) $J_{RR}/\Delta=0.05, 0.8, 3.2, 11.9$. The curves are off-set for clarity. Top inset: Sketch of the S/QD/S cotunnel junction consisting of a spinful quantum dot tunnel-coupled to the superconducting leads. Bottom inset: Second order self-energy diagram included in the calculation.}
\label{figintro}
\end{figure}

In this Letter, we perform a theoretical study of spin-induced YSR bound states and their influence on the nonequilibrium transport through S/QD/S cotunnel junctions. This study is motivated by recent experiments on quantum dots where the gate voltage can alter the nature of the quantum "impurity" as shown in Fig.\ref{figintro}; at specific odd fillings (3, 7, 11), the conductance through this carbon nanotube quantum dot is strikingly different from all the even-occupied fillings. The unusual behavior includes both negative differential conductance (NDC) and re-positioned conductance peaks inside the bias voltage range of $\pm 2\Delta/e$. Our modeling consists of a general framework for the nonequilibrium current through an interacting dot which is treated to second order in the lead-dot tunneling. Our treatment, therefore, does not include Kondo correlations which are, as we show below, not responsible for the unusual conductance features in the experimental plot in Fig.\ref{figintro}. Rather, as shown in the lower panel in Fig.\ref{figintro}, the low-bias Andreev transport is significantly altered by the presence of magnetic bound states which, in the asymmetric case, completely dominate the transport and generate NDC and shifted sub-gap conductance peaks similar to the experimental data. Specifically, the calculated IV curves in Fig.\ref{figintro} are obtained for fixed (varying) coupling to the left (right) lead. The bottommost curve is for a symmetric junction whereas the top curve is in the tunneling limit with highly asymmetric coupling. In the parameter regime between these two limiting cases, one clearly observes a transition from "conventional" MAR transport to a new double-step IV curve with associated NDC in the asymmetric limit. In addition, a regime of intermediate coupling asymmetry exists where mainly the current near $2\Delta$ is altered (see e.g. the red curve) reminiscent of fillings 7 and 11 in the experimental conductance map of Fig.\ref{figintro}. The absence of NDC at odd fillings 5 and 9 can be explained by a more symmetric coupling of the lower-lying nanotube orbital to the leads\cite{Grove:2009}.

We consider a system of two superconducting leads bridged by a quantum dot with its highest partially occupied orbital represented by a single-orbital Anderson model. By focussing on the cotunneling regime well inside a Coulomb diamond, where charge fluctuations happen only virtually, a Schrieffer-Wolff transformation\cite{Salomaa88} results in the following effective Hamiltonian
\begin{eqnarray}\nonumber
H&=&\frac{1}{2}\sum_{\alpha k}\psi_{\alpha k}^{\dagger}(\xi_{\alpha k}%
m^{3}+\Delta_{\alpha}^{^{\prime\prime}}m^{1}+\Delta_{\alpha}^{^{\prime}%
}m^{2})\psi_{\alpha k}\\
&+&\frac{1}{4}\sum_{\alpha^{\prime}k^{\prime},\alpha
k,i}J_{\alpha^{\prime}\alpha}S^{i}\psi_{\alpha^{\prime}k^{\prime}}^{\dagger
}m^{i}\psi_{\alpha k}.
\end{eqnarray}
Here, $\psi_{\alpha k}^{\dagger}=\left(
c_{\alpha k\uparrow}^{\dagger},
c_{\alpha -k\uparrow},
c_{\alpha k\downarrow}^{\dagger},
c_{\alpha -k\downarrow}
\right)$, $\xi_{\alpha k}$ is the quasiparticle energy, $\Delta_{\alpha}=\Delta^{\prime}_{\alpha}+i\Delta^{\prime\prime}_{\alpha}$ is the BCS-gap in lead $\alpha=L,R$, and the Nambu matrices are defined by $m^1=\tau^{1}\otimes\tau^{2}$, $m^{2}=\tau^{2}\otimes\tau^{2}$, $m^{3} =\tau^{3}\otimes\tau^{0}$, $m^x=\tau^3 \otimes \tau^1$, $m^y=\tau^0 \otimes \tau^2$, and $m^z=\tau^3 \otimes \tau^3$, where $\tau^{i}$ denote the Pauli matrices and $\tau^0$ is the $2\times 2$ identity matrix. $S^i$ is the quantum spin on the dot and $J_{\alpha^{\prime}\alpha}$ is the exchange-cotunneling amplitude with lead index $\alpha,\alpha'=L,R$, which satisfies the relation $J_{LR}J_{RL}=J_{LL}J_{RR}$.

The applied bias voltage, included through $H_{V}=-\sum_\alpha \mu_{\alpha}N_{\alpha}$ with $\mu_\alpha=\pm V/2$, may be incorporated into the spinors ${\psi}_{\alpha k,\eta}^{\dagger}(t)=\widetilde{\psi}_{\alpha k,\eta}^{\dagger}(t)e^{-it\mu_{\alpha}m_{\eta\eta}^{3}}$ ($\hbar=1$), where the time-dependence of $\widetilde{\psi}_{\alpha k,\eta}^{\dagger}(t)$ is now determined by the unbiased (non-interacting) leads. The corresponding contour ordered Green function transforms as $G_{\alpha\eta,\alpha^{\prime}\eta^{\prime}}(t,t^{\prime}) = e^{it\mu_{\alpha}m_{\eta\eta}^{3}}\widetilde{G}_{\alpha\eta,\alpha
^{\prime}\eta^{\prime}}(t,t^{\prime})e^{-it^{\prime}\mu_{\alpha^{\prime} }m_{\eta^{\prime}\eta^{\prime}}^{3}}$, and $\widetilde{G}$ can now be determined from perturbation theory in the transformed cotunneling term:
\begin{align}
H(t)  =\frac{1}{4}\sum_{\alpha^{\prime}k^{\prime}\eta^{\prime},\alpha k\eta
,i} \!\!\!\! J_{\alpha^{\prime}\alpha}^{\eta^{\prime}\eta}(t)S^{i}\widetilde{\psi
}_{\alpha^{\prime}k^{\prime}\eta^{\prime}}^{\dagger}(t)m_{\eta^{\prime}\eta
}^{i}\widetilde{\psi}_{\alpha k\eta}(t), \label{Hperp}
\end{align}
with the {\it time-dependent} amplitudes $J_{\alpha^{\prime}\alpha}^{\eta^{\prime}\eta}(t)=e^{-it\mu_{\alpha^{\prime}%
}m_{\eta^{\prime}\eta^{\prime}}^{3}}J_{\alpha^{\prime}\alpha}
e^{it\mu_{\alpha}m_{\eta\eta}^{3}}.$
The problem of voltage-biased superconducting junctions is inherently time-dependent and the Green functions acquire an explicit dependence on two times, $t$ and $t^\prime$. However, because of the harmonic dependence of the perturbation, the Green functions $G^{R,A,>,<}$ can be expressed as a Fourier series in $T=t+t'$
\begin{align}
G(t,t^{\prime})  &  =G(T,\tau)=\sum_{n}e^{inTV/2}G_{n}(\tau),
\end{align}
where $\tau=t-t^\prime$ is the relative time and $G_{n}(\omega)=\int_{-\infty}^{\infty}d\tau e^{i\omega\tau}G_{n}(\tau)$, with $G$ being shorthand for any of the functions $G^{<,>,R,A}$.

In the following, the Green functions and the self-energy are written as $8\times 8$ matrices containing $\alpha=L,R$ and the four Nambu indices $\eta=1,...,4$. Notice that the redundancy in the 4-spinor notation gives rise to anomalous Wick-contractions. Being identical replicas, however, these are effectively included by multiplying all self-energy diagrams by a factor of 4 and then employing the normal Dyson equation written here for $\widetilde{G}^R_{n}(\omega)$:
\begin{align}
& \widetilde{G}_{n}^{R}(\omega)=\delta_{n0}\widetilde{G}^{R(0)}(\omega
)+\widetilde{G}^{R(0)}(\omega-nV/2)\\
& \times \sum_{m}\widetilde{\Sigma}_{n-m}^{R}
(\omega-mV/2)\widetilde{G}_{m}^{R}(\omega+(n-m)V/2).\nonumber\label{DysonEq}
\end{align}
The solution generally takes the form
\begin{equation}
\widetilde{G}_{n}^{R}(\omega-nV/2)=[(1-\widetilde{M}^{R}(\omega))^{-1}]_{n0} \widetilde{G}^{R(0)}(\omega),\label{gDyson}
\end{equation}
where $\widetilde{M}_{nm}^{R}(\omega)=\widetilde{G}^{R(0)}(\omega-nV)\widetilde
{\Sigma}_{n-m}^{R}(\omega-(m+n)V/2)$ and the free ($k$-summed) Green function is given by
\begin{align}
\widetilde{G}^{R(0)}(\omega) =-\pi\nu_{F}\frac{\omega m^{0}+\Delta^{^{\prime\prime}}m^{1}+\Delta^{^{\prime}
}m^{2}}{\sqrt{|\Delta|^{2}-(\omega+i0_{+})^{2}}},\label{G0}
\end{align}
using a constant density of states, $\nu_{F}=1/(2D)$, with $D$ denoting the bandwidth. Using the irreducible self-energy, calculated to leading order in perturbation theory, Eq.(\ref{gDyson}) can be solved by numerical matrix inversion for each frequency $\omega$ after suitable truncation of the number of included harmonics $n$. Having done this matrix inversion, one obtains the $T$-matrix components
\begin{align}
\widetilde{T}_{n}^{R}(\omega) \! =  \! \sum_{m}\widetilde{\Sigma}_{n-m}^{R}(\omega-mV/2)[(1 \!- \!\widetilde
{M}(\omega+nV/2))^{-1}]_{m0},\label{Tmat}
\end{align}
\begin{align}\nonumber
& \widetilde{T}_{n}^{<}(\omega-n V/2) =\sum_{m}
\left[\widetilde{\Sigma}^{<}_{m}(\omega)+\widetilde{T}^{R}_{m}(\omega)
\left[\widetilde{G}^{(0)R}(\omega)\widetilde{\Sigma}^{<}_{m}(\omega) \right.\right. \\
& + \left. \left. \widetilde{G}^{(0)<}(\omega)\widetilde{\Sigma}^{A}(\omega)_{m}\right]
\right] [(1-\widetilde{M}(\omega+nV/2))^{-1}]_{mn},\label{TmatLess}
\end{align}
which enter the time-averaged current as
\begin{align}
\langle I \rangle =-e\int\!d\omega \sum_{\eta \eta'}\left\{\widetilde{T}_{L\eta,L\eta';0}(\omega)
\widetilde{G}^{(0)}_{L\eta',L\eta}(\omega)\right\}^{<}m_{\eta\eta}^{3}. \label{Curr}
\end{align}
This expression follows from the underlying Anderson model, and due to the energy dependence of the BCS density of states both $\widetilde{T}^{R}$ and $\widetilde{T}^{<}$ must be computed in order to determine the current. Given an expression for the conduction electron self-energy, this current formula applies to any dot Hamiltonian. The numerical cost lies in the matrix inversion, which becomes increasingly demanding as bias voltage is reduced, and a larger number of harmonics must be included to ensure convergence.
\begin{figure}[t]
\centering
\includegraphics[width=\columnwidth]{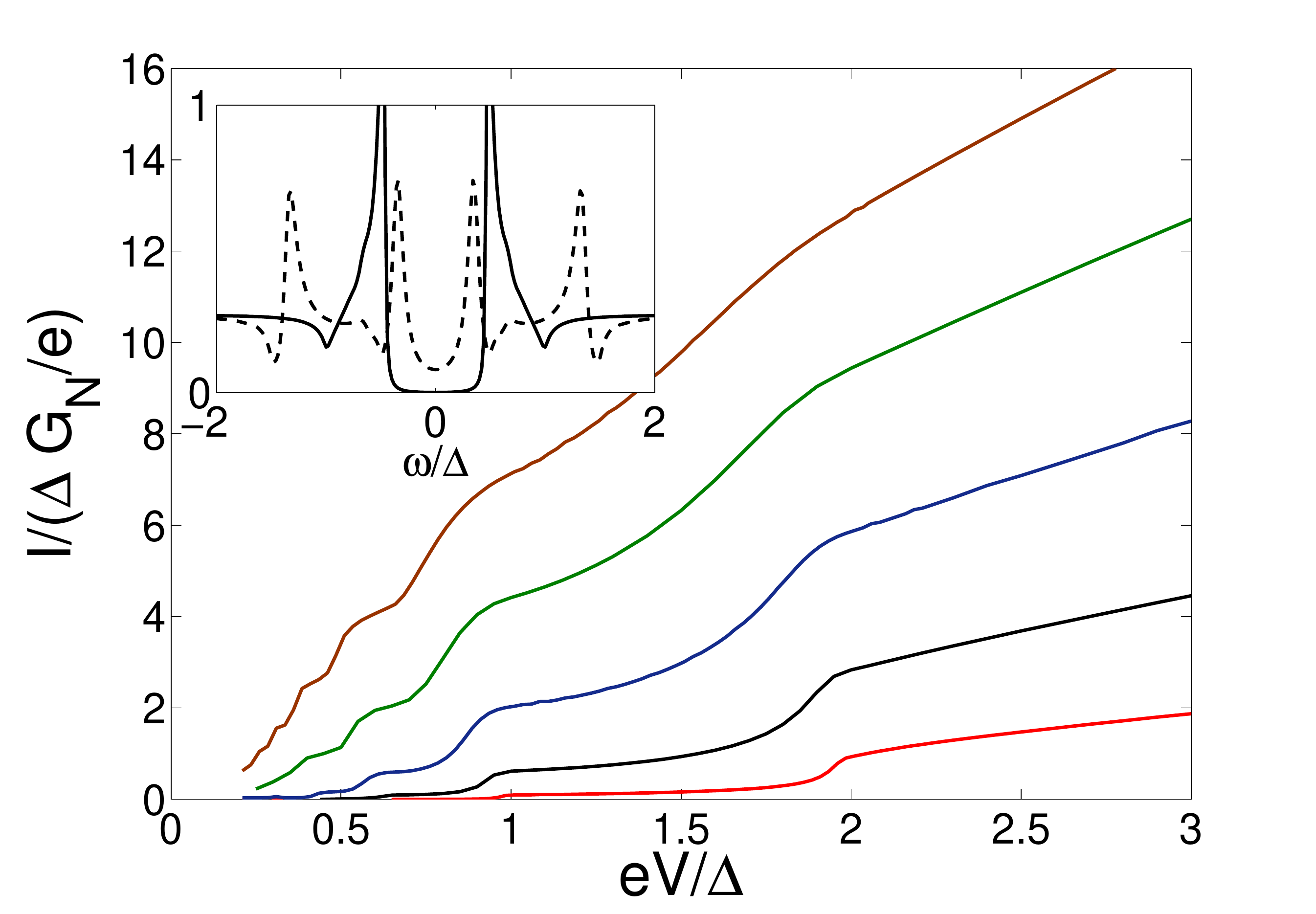}
\caption{(Color online) IV curves for a S/QD/S cotunnel junction with symmetric couplings, $J_{LL}=J_{RR}=J_{LR}=J$. From top to bottom the curves correspond to $J/\Delta=22.6, 17.3, 12.7, 8.8, 5.7$. All curves are normalized by $G_N$ determined for $J/\Delta=22.6$. Inset: dot spectral function for $J/\Delta=12.7$ with $V=0$ (solid) and $V=\Delta/e$ (dashed).}
\label{IVsym}
\end{figure}

\begin{figure}[t]
\centering
\includegraphics[width=\columnwidth]{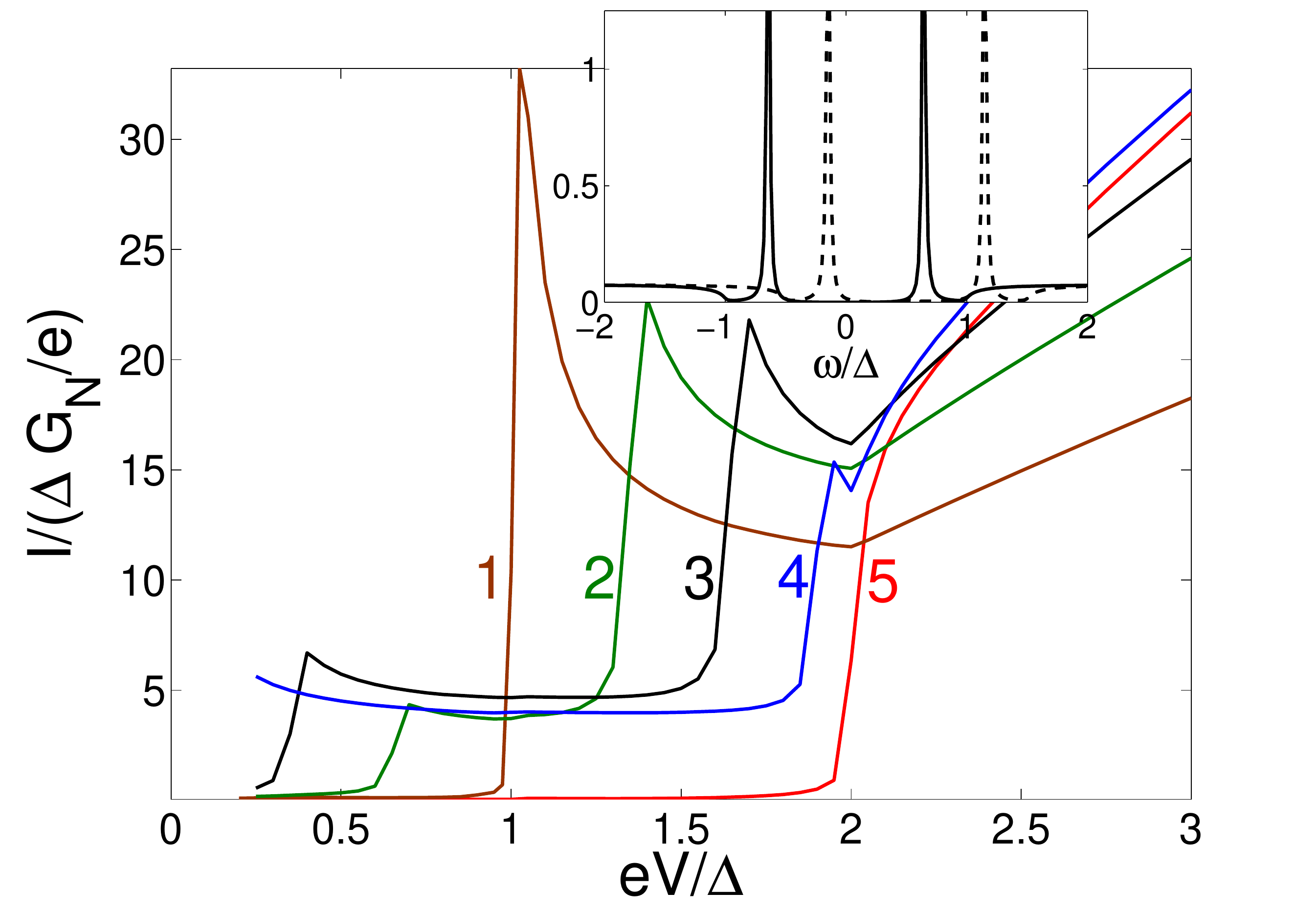}
\caption{(Color online) IV curves for a S/QD/S cotunnel junction with
coupling asymmetry $J_{RR}/J_{LL}=0.004,0.007,0.016,0.062,1.0$ for lines 1-5, respectively. All curves have $G_{N}=0.013 (2e^{2}/h)$, corresponding to $\nu_{F}J_{LL}\approx 0.744, 0.511, 0.335, 0.170, 0.042$.
Inset: as in Fig.\ref{IVsym} for the case with $J_{RR}/J_{LL}=0.016$.}
  \label{IVasym}
\end{figure}

For even-occupied (spinless) dots, the first order self-energy is exact and formulas (\ref{G0}-\ref{Curr}) quantitatively reproduce the IV curves presented, e.g. in Refs.~\cite{Averin95,Bratus:1995,Cuevas96}. Henceforth, we focus our attention to odd occupied dots where the spin turns out to play an important role. We restrict the discussion to the weak-coupling regime, $\nu_{F}J\lesssim 1$ ($J\equiv\max\{J_{\alpha'\alpha}\}$), and retain only the second order self-energy (see diagram in the lower inset of Fig.\ref{figintro}):
\begin{align}
& \widetilde{\Sigma}_{\alpha^{\prime}\alpha,\gamma^{\prime}\gamma,n}
^{R(2)}(\omega)=\frac{1}{16}\sum_{\eta^{\prime}\eta
,i,\alpha^{\prime\prime}}J_{\alpha^{\prime}\alpha^{\prime\prime}}
J_{\alpha^{\prime\prime}\alpha}m_{\gamma^{\prime}\eta^{\prime}}^{i} m_{\eta\gamma}^{i} \\\nonumber
& \times \widetilde{G}_{\eta\eta^\prime}^{R(0)}(\omega-[\mu_{\alpha^{\prime}} m_{\gamma^{\prime
}\gamma^{\prime}}^{3}-\mu_{\alpha
^{\prime\prime}}m_{\eta^{\prime}\eta^{\prime}}^{3}+ nV/2]),
\end{align}
together with the corresponding advanced and lesser components.
For all numerical evaluations presented in Figs.~\ref{figintro}-\ref{IVasym}, we take $D=20\Delta$.
 
For equal coupling to the two leads, the IV curves for the spinful dot are shown in Fig.\ref{IVsym}. Evidently, the current is dominated by MAR steps at $2\Delta/n$, resembling that of an even occupied dot, and there appears to be little effect of the spin. This is in stark contrast to the case where tunnel couplings to the two leads are very different. As seen in Fig.\ref{IVasym}, the coupling asymmetry significantly enhances the sub-gap current and the IV characteristics now exhibit a marked cusped two-step shape. We can understand this evolution from symmetric to asymmetric junctions by taking a closer look at the physical retarded $T$-matrix, obtained as:
\begin{equation}
T_{\alpha\eta,\alpha^{\prime}\eta^{\prime};n}^{R}(\omega)= \widetilde{T}_{\alpha\eta,\alpha^{\prime}\eta^{\prime};n-(\mu_{\alpha} m_{\eta\eta}^{3}-\mu_{\alpha^{\prime}} m_{\eta^{\prime}\eta^{\prime}}^{3})/V}^{R}(\omega^\prime),
\end{equation}
where  $\omega^\prime=\omega+(\mu_{\alpha}m_{\eta\eta}^{3}+\mu_{\alpha^{\prime}} m_{\eta^{\prime}\eta^{\prime}}^{3})/2$. In the insets of Figs. \ref{IVsym}-\ref{IVasym} we plot $-\mbox{Im}\widetilde{T}_{L1,L1;0}^{R}(\omega+V/2)/\pi$, which is proportional to the {\it time-averaged} dot-electron spectral function for the underlying Anderson model, for respectively $V=0$ and $V=\Delta/e$. In the symmetric limit (Fig.\ref{IVsym}) sub-gap features are predominantly due to MAR, and for $V\neq 0$ these disperse strongly with time and are therefore smeared out by time averaging. This results in broad resonances which depend significantly on the bias voltage. In the asymmetric limit (Fig.\ref{IVasym}) the spectrum shows a pair of YSR Andreev bound states~\cite{Balatsky:2006, Koerting:2010} located symmetrically around the left chemical potential at $\mu_{L}\pm\omega_{S}$ where $\omega_{S}=\Delta(1-x)/(1+x)$ with $x=3(\pi\nu_{F}J_{LL}/4)^{2}$ and hence crossing each other when $\nu_{F}J_{LL}=4/(\pi\sqrt{3})\approx 0.74$, corresponding to curve 1 in Fig.~\ref{IVasym}. Note that the peaks in the inset have been smeared with an artificial broadening chosen small enough that it does not affect the calculated current. In contrast to the symmetric case, these bound states do not disperse with time and hence survive the time-averaging. This is seen explicitly in the inset of Fig.~\ref{IVasym} where peaks at $\omega_{S}=0.66\Delta$ simply shift by $eV/2$ upon an applied bias voltage. The values of $\omega_{S}$ are directly related to the cusps in the current at $eV=\Delta\pm\omega_{S}(=1.66, 0.34)$ (case 3 in the main panel of Fig.\ref{IVasym}), corresponding to the matching up of YSR bound states with the coherence peaks of the weakly probing right lead. In this case, the current is dominated by Andreev processes matching up with the symmetrically positioned bound states. This is similar to the resonant MAR discussed for noninteracting junctions\cite{LeviYeyati:1997, Johansson:1999} but now the resonant effect is enhanced because the bound states are positioned symmetrically around $\mu_{L}$, and therefore match both an incoming electron {\it and} an outgoing hole of a contributing Andreev reflection process. Finally, a strong coupling asymmetry is seen to change the quasiparticle tunneling step at $V=2\Delta/e$ into a local current minimum, consistent with the experimental observations of NDC in Refs.~\cite{Eichler:2007, Jespersen:2007, Grove:2009,Pascual:2011}, and seen also in Fig.\ref{figintro}.

The progression of curves in Fig.\ref{IVasym} corresponds to increasing asymmetry factor $J_{LL}/J_{RR}$ with fixed normal state conductance, $G_N=(3\pi^2/4)|\nu_{F}J_{LR}|^2 (2e^2/h)$ (calculated to second order), and therefore also to increasing values of $J_{LL}$. For sufficiently weak couplings, the dominant higher order self-energy corrections may be included by using the renormalized exchange coupling obtained from Poor man's scaling from $D$ down to $\Delta$~\cite{Anderson:1970}. Thus replacing $\nu_{F}J_{LL}$ with $1/\ln(\Delta/T_{K})$, we conclude that the two YSR bound states lie at zero energy when $1/\ln(\Delta/T_{K})\approx 0.74$, i.e. $T_{K}\approx 0.26\Delta$. This is consistent with numerical renormalization group calculations for the equilibrium problem\cite{Satori:1992, Yoshioka:2000, Bauer:2007}, which show a transition from doublet to singlet groundstate, corresponding to a crossing of the bound states, at $T_{K}\sim 0.3\Delta$. For even stronger couplings, the YSR bound states again lie symmetrically around $\mu_{L}$ and our calculations would give IV curves which look qualitatively similar to those in Fig.\ref{IVasym}. For such couplings, however, the Kondo screening is so strong that we can no longer rely on our perturbative approach. Note, however, that our approach would be exact in all regimes if dealing with a classical magnetic impurity, potentially relevant for experiments probing impurities on surfaces~\cite{Moca:2008,Pascual:2011}. 

In summary, we have presented a general Hamiltonian approach to the nonequilibrium current through an interacting quantum dot contacted by superconducting leads. For odd-occupied dots, the spin leads to IV characteristics being strongly dependent on coupling asymmetry, giving rise to negative differential conductance and shifted sub-gap conductance peaks in agreement with experiments. More work is required to study the asymmetry variations in the cross-over regime towards stronger couplings, $T_{K}>\Delta$, where Kondo screening eventually quenches local superconducting fluctuations.

B.M.A. acknowledges support from The Danish Council for Independent Research $|$ Natural Sciences.


\begin{thebibliography}{100}
\bibitem{Sheer:1997} E. Scheer {\it et al.}, Phys. Rev. Lett. {\bf 78}, 3535 (1997).
\bibitem{Buitelaar:2002} M. R. Buitelaar, T. Nussbaumer, and C. Sch{\"o}nenberger, Phys. Rev. Lett. {\bf 89}, 256801 (2002).
\bibitem{Buitelaar:2003} M. R. Buitelaar {\it et al.}, Phys. Rev. Lett. {\bf 91}, 057005 (2003).
\bibitem{Kasumov:2005} A. Yu. Kasumov {\it et al.}, Phys.~Rev.~B \textbf{72}, 033414 (2005).
\bibitem{JarilloHerrero:2006} P. Jarillo-Herrero, J. A. van Dam, and L. P. Kouwenhoven, Nature {\bf 439}, 953 (2006).
\bibitem{Heersche:2006} H. B. Heersche {\it et al.}, Nature \textbf{446}, 56 (2007).
\bibitem{Xiang:2006} J. Xiang {\it et al.}, Nature Nanotech. {\bf 1}, 208 (2006).
\bibitem{Averin95}D.~Averin and A.~Bardas, Phys.~Rev.~Lett.~\textbf{75}, 1831 (1995).
\bibitem{Bratus:1995} E. N. Bratus, V. S. Shumeiko, and G. Wendin, Phys. Rev. Lett. {\bf 74}, 2110 (1995).
\bibitem{Cuevas96}J.~C.~Cuevas, A.~Martin-Rodero, and A.~Levy Yeyati,
Phys.~Rev.~B \textbf{54}, 7366 (1996).
\bibitem{Eichler:2007} A.~Eichler {\it et al.}, Phys.~Rev.~Lett. \textbf{99}, 126602 (2007).
\bibitem{Jespersen:2007}
T.~S.~Jespersen {\it et al.}, Phys.~Rev.~Lett. \textbf{99}, 126603 (2007).
\bibitem{Buizert:2007} C. Buizert {\it et al.}, Phys.~Rev.~Lett. \textbf{99}, 136806 (2007).
\bibitem{Grove:2009} K.~Grove-Rasmussen {\it et al.}, Phys.~Rev.~B \textbf{79}, 134518 (2009).
\bibitem{Pillet:2010}J.~D.~Pillet, {\it et al.}, Nat. Phys. {\bf 6}, 965 (2010).
\bibitem{Balatsky:2006} For a review see, A. V. Balatsky, I. Vekhter, and J.-X. Zhu, Rev. Mod. Phys. {\bf 78}, 373 (2006).
\bibitem{Moca:2008}C.~P.~Moca {\it et al.}, Phys. Rev. B \textbf{77}, 174516 (2008).
\bibitem{Pascual:2011}K. J. Franke, G. Schulze, and J. I. Pascual, Science \textbf{332}, 940 (2011).
\bibitem{Glazman:1989}L.~Glazman and M. Matveev, JETP Letters {\bf 49}, 659 (1989).
\bibitem{Satori:1992}K.~Satori {\it et al.}, J. Phys. Soc. Jpn. \textbf{61}, 3239 (1992).
\bibitem{Rozhkov:1999} A. V. Rozhkov and D. P. Arovas, Phys. Rev. Lett. {\bf 82}, 2788 (1999).
\bibitem{Clerk:2000} A. A. Clerk and V. Ambegaokar, Phys. Rev. B \textbf{61}, 9109 (2000).
\bibitem{Yoshioka:2000}T.~Yoshioka and Y.~Ohashi, J. Phys. Soc. Jpn. \textbf{69}, 1812 (2000).
\bibitem{Vecino:2003} E. Vecino, A. Martin-Rodero, and A. Levy Yeyati, Phys. Rev. B {\bf 68}, 035105 (2003).
\bibitem{Choi:2004} M.-S. Choi {\it et al.}, Phys. Rev. B {\bf 70}, 020502(R) (2004).
\bibitem{Siano:2004} F. Siano and R. Egger, Phys. Rev. Lett. {\bf 93}, 047002 (2004).
\bibitem{Oguri:2004} A. Oguri, Y. Tanaka, and A. C. Hewson, J. Phys. Soc. Jpn. {\bf 73}, 2494 (2004).
\bibitem{Sellier:2005}G.~Sellier {\it et al.}, Phys. Rev. B \textbf{72}, 174502 (2005).
\bibitem{Bauer:2007}J. Bauer, A. Oguri, and A. C. Hewson, J. Phys.: Cond. Mat. \textbf{19}, 486211 (2007).
\bibitem{Karrasch:2008} C. Karrasch, A. Oguri, and V. Meden, Phys. Rev. B \textbf{77}, 024517 (2008).
\bibitem{Hecht:2008}T. Hecht {\it et al.}, J. Phys.: Cond. Mat. \textbf{20}, 275213 (2008).
\bibitem{Meng:2009} T. Meng, S. Florens, and P. Simon, Phys. Rev. B \textbf{79}, 224521 (2009).
\bibitem{Avishai:2001} Y. Avishai, A. Golub, and A. D. Zaikin, Phys. Rev. B {\bf 63}, 123415 (2001); {\it ibid.}  {\bf 67}, 041301 (2003).
\bibitem{LevyYeyati:2003} A. Levy Yeyati, A. Martin-Rodero, and E. Vecino, Phys. Rev. Lett. {\bf 91}, 266802 (2003).
\bibitem{Liu:2004} S. Y. Liu and X. L. Lei, Phys. Rev. B {\bf 70}, 205339 (2004).
\bibitem{DellAnna:2008} L. Dell'Anna, A. Zarunov, and R. Egger, Phys. Rev. B {\bf 77}, 104525 (2008).
\bibitem{Salomaa88}M.~M.~Salomaa, Phys. Rev. B \textbf{37}, 9312 (1988).
\bibitem{Koerting:2010} V.~Koerting {\it et al.}, Phys.~Rev.~B \textbf{82}, 245108 (2010).
\bibitem{LeviYeyati:1997} A. Levy Yeyati {\it et al.}, Phys. Rev. B {\bf 55}, R6137 (1997).
\bibitem{Johansson:1999} G. Johansson {\it et al.}, Phys. Rev. B {\bf 60}, 1382 (1999).
\bibitem{Anderson:1970}P.~W.~Anderson, J. Phys. C {\bf 3}, 2436 (1966).
\end{thebibliography}
\end{document}